 \documentclass[final,1p,times]{elsarticle}
 
\usepackage{amssymb}
\usepackage{subfigure,dcolumn}
\usepackage{epstopdf}
\usepackage{graphicx}
\usepackage{float}
\usepackage{subfigure}
\usepackage{url}
\usepackage{amsmath}
\usepackage{multirow}
\usepackage{ulem}
\usepackage{siunitx}
\usepackage{blindtext}
\usepackage{bm}
\usepackage{lineno}           
\biboptions{sort&compress}

\newcommand{\Lat}{L_{\rm Att}}
\newcommand{\Tstart}{T_{\rm start}}
\newcommand{\Tstop}{T_{\rm stop}}

\newcommand{\kl}{K_{L}}
\newcommand{\EE}{e^+e^-}

\newcommand{\jpsi}{J/\psi}

\newcommand{\bcntr}{\begin{center}}
\newcommand{\ecntr}{\end{center}}
\newcommand{\beq}{\begin{equation}}
\newcommand{\eeq}{\end{equation}}
\newcommand{\beqar}{\begin{eqnarray}}
\newcommand{\eeqar}{\end{eqnarray}}
\newcommand{\bitm}{\begin{itemize}}
\newcommand{\benu}{\begin{enumerate}}

\newcommand{\bitmb}{\begin{itemize}}
\newcommand{\benub}{\begin{enumerate}}
\newcommand{\eitm}{\end{itemize}}

\newcommand{\bfrm}{\begin{frame}}
\newcommand{\efrm}{\end{frame}}
\newcommand{\bct}{\begin{center}}
\newcommand{\ect}{\end{center}}
\newcommand{\bclm}{\begin{columns}}
\newcommand{\eclm}{\end{columns}}
\newcommand{\bpic}{\begin{overpic}}
\newcommand{\epic}{\end{overpic}}
\newcommand{\bblk}{\begin{block}}
\newcommand{\eblk}{\end{block}}

\newcommand{\eenu}{\end{enumerate}}

\newcommand{\ps}{\si{\pico\second}}
\newcommand{\s}{\si{\second }}

\newcommand{\m}{\si{\metre}}

\newcommand{\mm}{\si{\milli\metre}}
\newcommand{\cm}{\si{\centi\metre}}

\newcommand{\V}{\si{\volt}}
\newcommand{\mV}{\si{\milli\volt}}

\newcommand{\ns}{\si{\nano\second}}

\newcommand{\ohm}{\si{\ohm}}

\newcommand{\mhz}{\si{\mega\Hz}}
\newcommand{\ghz}{\si{\giga\Hz}}

\newcommand{\Trise}{\tau_{\rm rise}}

\newcommand{\gevc}{\hbox{GeV}/c}
\newcommand{\gev}{\hbox{GeV}}

\def\Journal#1#2#3#4{{#1} {\bf #2} (#4) #3}

\def\NIMA{Nucl. Instrum. Methods A}

\def\PRD{Phys. Rev. D}

\def\PTEP{Prog. Theor. Exp. Phys.}

\def\NST{Nucl. Sci. Tech.}
\def\JINST{J. Instrum.}
\def\TNS{IEEE Trans. Nucl. Sci.}

\journal{\NIMA}

\begin{document}

\begin{frontmatter}

\title{R\&D of KLM Upgrade for Direct Measurement of Neutral Hadron Momentum via Time-of-Flight in Belle II} 
%

\author[aff1]{Xiyang Wang}

\author[aff1]{Hongyu Zhang}

\author[aff1]{Shiming Zou}

\author[aff3]{Zibing Bai}

\author[aff1,aff2]{Deqing Fang\corref{cor1}}
\ead{dqfang@fudan.edu.cn}

\author[aff1]{Kairui Huang}

\author[aff3]{Ziyu Liu}

\author[aff1,aff2]{Yugang Ma\corref{cor1}}
\ead{mayugang@fudan.edu.cn}

\author[aff3]{Weiqi Meng}

\author[aff1]{Ting Wang}

\author[aff1]{Xiaolong Wang\corref{cor1}}
\ead{xiaolong@fudan.edu.cn}
\cortext[cor1]{Corresponding author}

\author[aff1]{Shiqing Xie}

\author[aff1]{Mingjie Yang}

\author[aff3]{Junhao Yin}

\author[aff1]{Mingkuan Yuan}

\author[aff1]{Wanyi Zhuang}

\affiliation[aff1]{organization={Key Laboratory of Nuclear Physics and Ion-beam Application (MOE) and Institute of Modern Physics, Fudan University},
            addressline={220 Handan Road},
            city={Shanghai},
            postcode={200433},
            country={China}}
            
\affiliation[aff2]{organization={Shanghai Research Center for Theoretical Nuclear Physics, NSFC and Fudan University},
            city={Shanghai},
            postcode={200438},
            country={China}}

\affiliation[aff3]{organization={School of Physics, Nankai University},
            addressline={94 Weijin Road, Nankai District},
            city={Tianjin},
            postcode={300071},
            country={China}}

\begin{abstract}
Accurate momentum determination of a neutral hadron, such as a $\kl$ meson or a neutron, remains a significant challenge in particle physics and nuclear physics experiments. The Belle II experiment presents an opportunity to address this challenge through an upgrade incorporating Time-of-Flight (TOF) capability for its large $\kl$ and Muon Detector (KLM). We investigate the feasibility of momentum determination via TOF measurement. To achieve high time resolution for the KLM upgrade, we conduct research and development of cost-effective plastic scintillators in collaboration with GaoNengKeDi Company, and technology utilizing silicon photomultipliers (SiPMs) arrays. A bulk attenuation length of $120 \pm 7~\cm$ has been achieved with a $135~\cm$-long sample, along with a time resolution of $70\pm 7~\ps$ at its midpoint.  A $50~\cm$-long scintillator demonstrates an exceptional time resolution of $47 \pm 2~\ps$. These results highlight the potential of the proposed technology for improving neutral hadron momentum measurements in an upgraded Belle II KLM detector.
\end{abstract}

\begin{keyword}
Time-of-Flight \sep Time resolution \sep Muon detector \sep Scintillator \sep SiPM
\end{keyword}

\end{frontmatter}


\section{Introduction}
\label{sec:I}

In particle and nuclear physics experiments, accurately determining the momentum of neutral hadrons, such as $\kl$ mesons and neutrons, poses significant challenges. High energy frontier experiments like ATLAS and CMS employ costly hadron calorimeters (HCALs) that achieve energy resolutions $\sigma(E)/E$ of approximately $(50 - 90)\%/\sqrt{E(\gev)}$~\cite{HCAL_ATLAS, HCAL_CMS}. On the other hand, intensity frontier experiments such as BaBar~\cite{BaBar}, Belle~\cite{Belle}, Belle II~\cite{belle2_det}, and BESIII~\cite{BESIII} do not use HCALs for measuring the energy of neutral hadrons with momenta ranging from $1 - 5~\gevc$. This is primarily due to the limited energy resolution that can be achieved by HCAL.

As a super B-factory experiment, the Belle II experiment was designed to explore new physics at the forefront of high-luminosity collisions and improve the precision of measurements for Standard Model parameters~\cite{belle2_phy}. The Belle II spectrometer has a huge outermost subdetector, the $\kl$ and Muon Detector (KLM)~\cite{belle2_det}. In the Conceptual Design Report (CDR) for the Belle II upgrade, we proposed enhancing the KLM system with a new capability: Time-of-Flight (TOF) measurements~\cite{B2U_CDR}. This upgrade will enable the direct determination of momentum with a resolution of approximately 10\% for a long-lived neutral hadron, such as a $\kl$ meson or a neutron, based on its velocity. However, implementing TOF technology poses a significant challenge due to the need for cost-effectiveness, given the large volume and number of readout channels of the KLM system. 

Traditional TOF detectors primarily employ two technological approaches. The first approach utilizes plastic scintillators coupled with photodetectors, such as photomultiplier tubes (PMTs) or silicon photomultipliers (SiPMs). The PMT method has been successfully implemented in TOF systems such as the Belle TOF with a time resolution of $100~\ps$~\cite{Belle} and the BESIII barrel TOF with a resolution of $80~\ps$ using two scintillator layers~\cite{BESIII}. PMTs have inherent limitations, including their large physical size, which creates dead zones, and the requirement for a high-voltage (HV) system operating at over $1000~\V$. The inherent transit time spread of PMTs poses a significant challenge in achieving high time resolution (HTR). In addition, the use of such scintillators and PMTs also incurs significant costs. The second approach utilizes multi-gap resistive plate chambers (MRPCs), as in experiments such as STAR~\cite{STAR}, ALICE~\cite{ALICE}, and the upgraded endcap TOF of BESIII~\cite{BESIII-ETOF}. While MRPCs can achieve an HTR of $50-60~\ps$, they suffer from notable drawbacks, including complex structures requiring gas flow systems, susceptibility to gas contamination (e.g., $\rm C_2H_2F_4/C_4H_{10}/SF_6$), and a large number of readout channels.

Currently, several experiments have successfully implemented SiPMs with small-sized scintillators (length $\le 15~\cm$) to construct TOF detectors, achieving time resolutions better than $100~\ps$, as demonstrated by the PANDA~\cite{PANDA-TOF}, AMS-100~\cite{AMS-100-TOF}, MEG-II~\cite{MEG-II,MEG-II2} and Mu3e~\cite{Mu3e} experiments. However, the application of SiPMs with large elongated scintillators remains limited. Key challenges in such implementations include the scintillator's attenuation length, elevated noise levels in SiPM arrays, and issues associated with slow rise times ($\Trise$) of the signal pulse.

To achieve HTR for the KLM upgrade~\cite{B2U_CDR}, the limitations of TOF technologies using PMTs or MRPCs underscore the need for alternative technologies and materials that balance performance, simplicity, and cost-effectiveness. SiPM-based technology presents a promising solution, though challenges related to large-scale applications must be addressed. In collaboration with us, the GaoNengKeDi Company~\cite{GNKD} has been developing new scintillators for HTR at low cost. Simultaneously, we have been conducting research and development (R\&D) on SiPM-based technology for HTR. Through these efforts, we have successfully developed new scintillators and SiPM-based TOF technology for HTR, which are critical for an upgraded KLM to determine the momentum of a long-lived neutral hadron. This paper describes the principle for measuring the momentum of a neutral hadron using an upgraded KLM with HTR, introduces the new scintillators and SiPM-based technologies developed for HTR, and demonstrates their performance.

\section{Belle II KLM and its upgrade}

\begin{figure}[htb]
\bcntr
\includegraphics[width=0.45\textwidth]{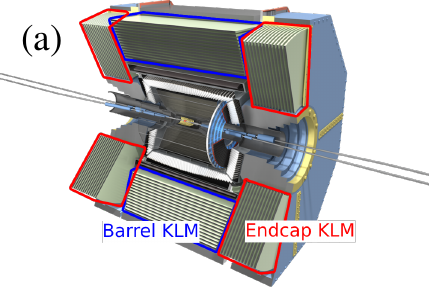}
\includegraphics[width=0.4\textwidth]{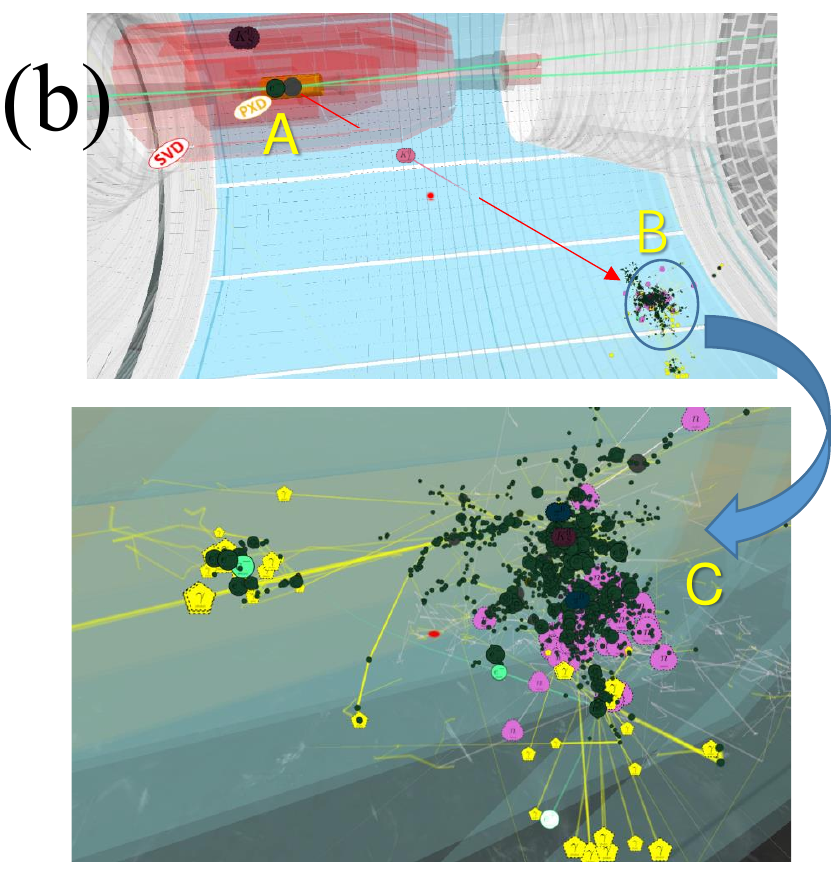}
\ecntr
\caption{(a) The KLM system within the Belle II detector. (b) A hadronic cluster generated by a $\kl$ meson in the Belle II virtual reality simulation. Point A illustrates the $\EE$ collision that produces the $\kl$ meson. Point B depicts the $\kl$ meson initiating a hadronic shower within the KLM system, while Point C provides a detailed visualization of the hadronic shower processes.}
\label{Belle2KLM}
\end{figure}

As the outermost component of the Belle II detector, the KLM is responsible for identifying $\kl$ mesons and muons with momenta up to $4.5~\gevc$~\cite{belle2_det}. As illustrated in Fig.~\ref{Belle2KLM}(a), the KLM spans radii from $200~\cm$ to $~240~\cm$ in the octagonal barrel region and from $130~\cm$ to $340~\cm$ in the endcaps. The barrel section consists of 15 detector layers interspersed with 14 layers of yoke iron, while each endcap contains 14 detector layers and 14 iron layers. The iron yoke serves a dual purpose: it acts as a magnetic flux return for the solenoid and enables hadronic shower, which is essential for $\kl$ detection.

The large-surface-area detector panels, approximately $3.1~\cm$ thick, are positioned between $4.7~\cm$ steel plates. Within these panels, the innermost two layers of the barrel and all layers of the endcaps are equipped with extruded scintillator detector modules~\cite{eklm}. The remaining 13 barrel layers utilize the Belle experiment's legacy Resistive Plate Chambers. In the scintillator modules, each scintillator strip incorporates a Kuraray Y11(200)MSJ wavelength-shifting (WLS) fiber, which collects photons emitted by the extruded scintillator and guides them to a Hamamatsu (HPK) MPPC S10362-13-050C, a SiPM with a sensitive area of $1.3~\mm \times 1.3~\mm$~\cite{HPK}.

In the CDR of the Belle II upgrade~\cite{B2U_CDR}, we proposed the idea of enhancing the KLM system with the capability to \textbf{measure the momenta of $\kl$ mesons through a TOF measurement using new scintillators}. The momentum $p$ of a neutral hadron with velocity $\beta$ is expressed as:
\beq\label{eq_1}
p = \gamma m v = \frac{mc\beta}{\sqrt{1-\beta^2}} = \frac{mcL}{\sqrt{T_{\rm fly}^2c^2 - L^2}},
\eeq
where $\gamma$ is the Lorentz factor, $m$, $L$, and $T_{\rm fly}$ are the nominal mass, the flight length, and the flight time of this hadron. Since $L/T_{\rm fly} \approx c$ in practical measurements, the relationship between the uncertainty in momentum ($\sigma_p$) and the uncertainty in flight time ($\sigma_{T_{\rm fly}}$) is given by:
\beq\label{eq_2}
\frac{\sigma_p}{p} = \frac{T_{\rm fly} p^2}{m^2 L^2} \cdot \sigma_{T_{\rm fly}} \approx 
\frac{\gamma^2c}{L}\cdot \sigma_{T_{\rm fly}}.
\eeq
Our Geant4-based simulations~\cite{Geant4} of an upgraded KLM indicate that measuring $T_{\rm fly}$ of a $\kl$ meson is achievable~\cite{B2U_CDR}. It relies on the precise measurements of the start time ($\Tstart$) and stop time ($\Tstop$).

The majority of $K_L$ mesons are produced directly at the interaction point (IP), and the $\Tstart$ can be determined from the collision time at the SuperKEKB accelerator~\cite{SuperKEKB}, which achieves a resolution better than $32~\ps$. For those originating from the decay of long-lived hadrons, such as $B$ or $D$ mesons, the lifetime of the parent meson is typically less than $10~\ps$~\cite{PDG}. Consequently, the expected resolution for $\Tstart$ is projected to be better than $35~\ps$. 

The $\Tstop$ is determined using timing information from multiple hits within the hadronic cluster detected by the upgraded KLM system. Figure~\ref{Belle2KLM}(b) illustrates a hadronic cluster generated by a $\kl$ meson, as visualized in the Belle II virtual reality simulation~\cite{B2_VR}. The simulation data reveal multiple hits within this hadronic cluster. Assuming a time resolution of $100~\ps$ for $\Tstop$ and a typical flight length of $L = 200~\cm$ in Belle II, the $\sigma_p$ for a $\kl$ meson with $p = 1.5~\gevc$ can achieve $0.2~\gevc$, corresponding to a relative uncertainty of 13\%. The upgraded KLM enhances the study of final states involving a $\kl$ meson, such as $B^0\to \jpsi\kl$, while simultaneously reducing accidental backgrounds. In scenarios where the momentum of a $\kl$ cannot be determined solely from kinematic constraints, as in the decay $B^+ \to K^+ \kl\kl$, the new KLM enables the experimental investigation of such processes.

\section{Improvements to HTR scintillation detector}
\label{sec:II}

The current scintillation detection scheme in the KLM achieves a time resolution of only $1-2~\ns$, as demonstrated in laboratory tests~\cite{KLMSC}. This level of performance is insufficient for $T_{\rm fly}$ measurement, highlighting the necessity for an upgraded detection scheme. As described in the Belle II Upgrade CDR~\cite{B2U_CDR}, several key improvements need to be implemented.

WLS fibers exhibit limited time resolution due to their low light collection efficiency, primarily caused by their small diameter, and the additional conversion process of fluorescent light. Additionally, extruded scintillators typically have an attenuation length of approximately $10~\cm$. It is difficult to use the combination of extruded scintillators and WLS fibers for HTR. Instead, we focus on developing new scintillators with much longer attenuation lengths. To enhance light collection efficiency, we employ a SiPM array, which forms a large sensitive surface for coupling with the new scintillator. Achieving the best possible time resolution requires the development of new front-end electronics and the optimization of SiPM array configurations. Furthermore, strategically mounting SiPM arrays at both ends of a scintillator strip has been shown to enhance time resolution significantly.

\subsection{Development of new HTR scintillators}

\begin{figure}[htb]
\centering
\includegraphics[width=0.45\textwidth]{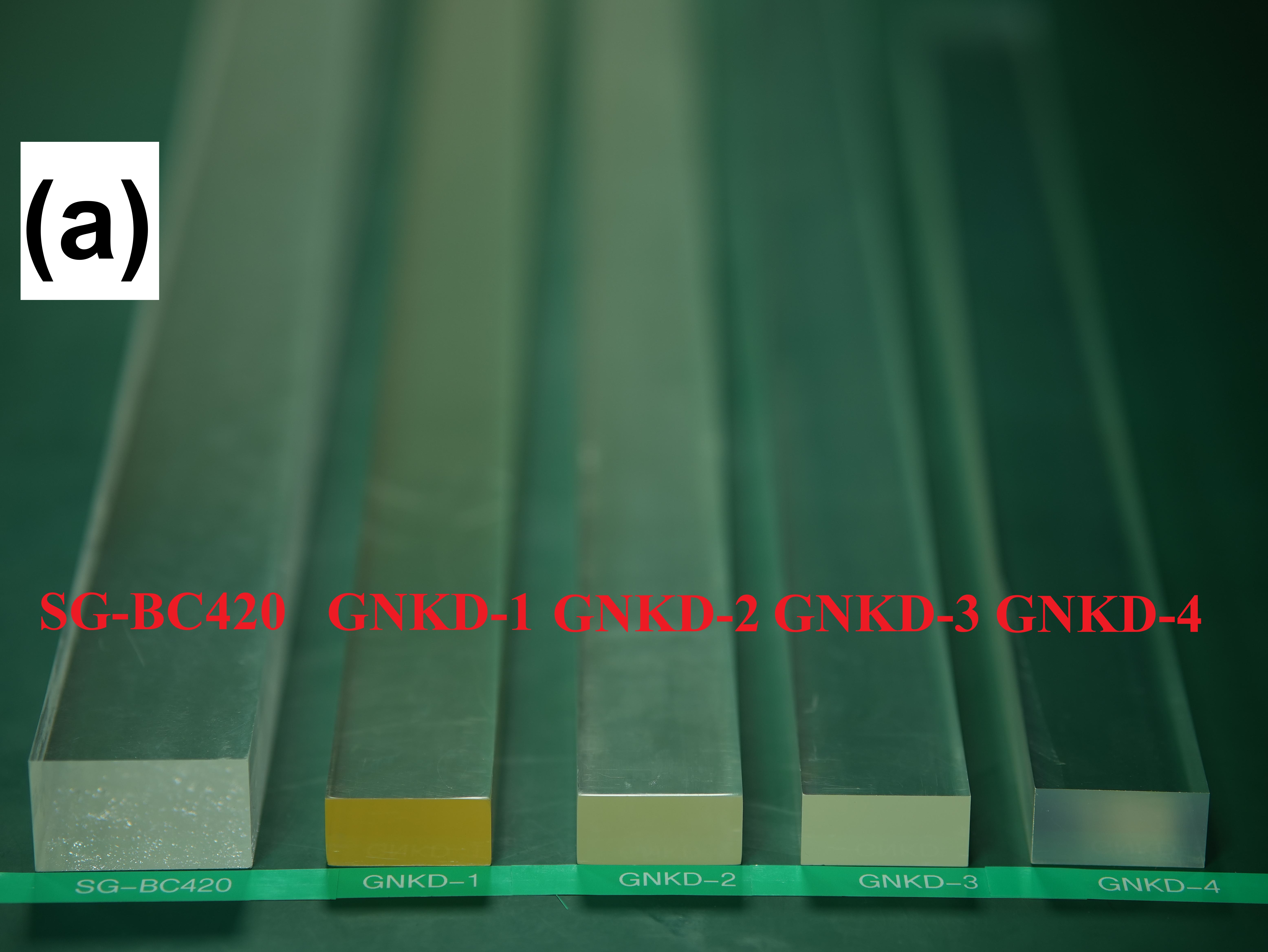}\\
~\\
\includegraphics[width=0.45\textwidth]{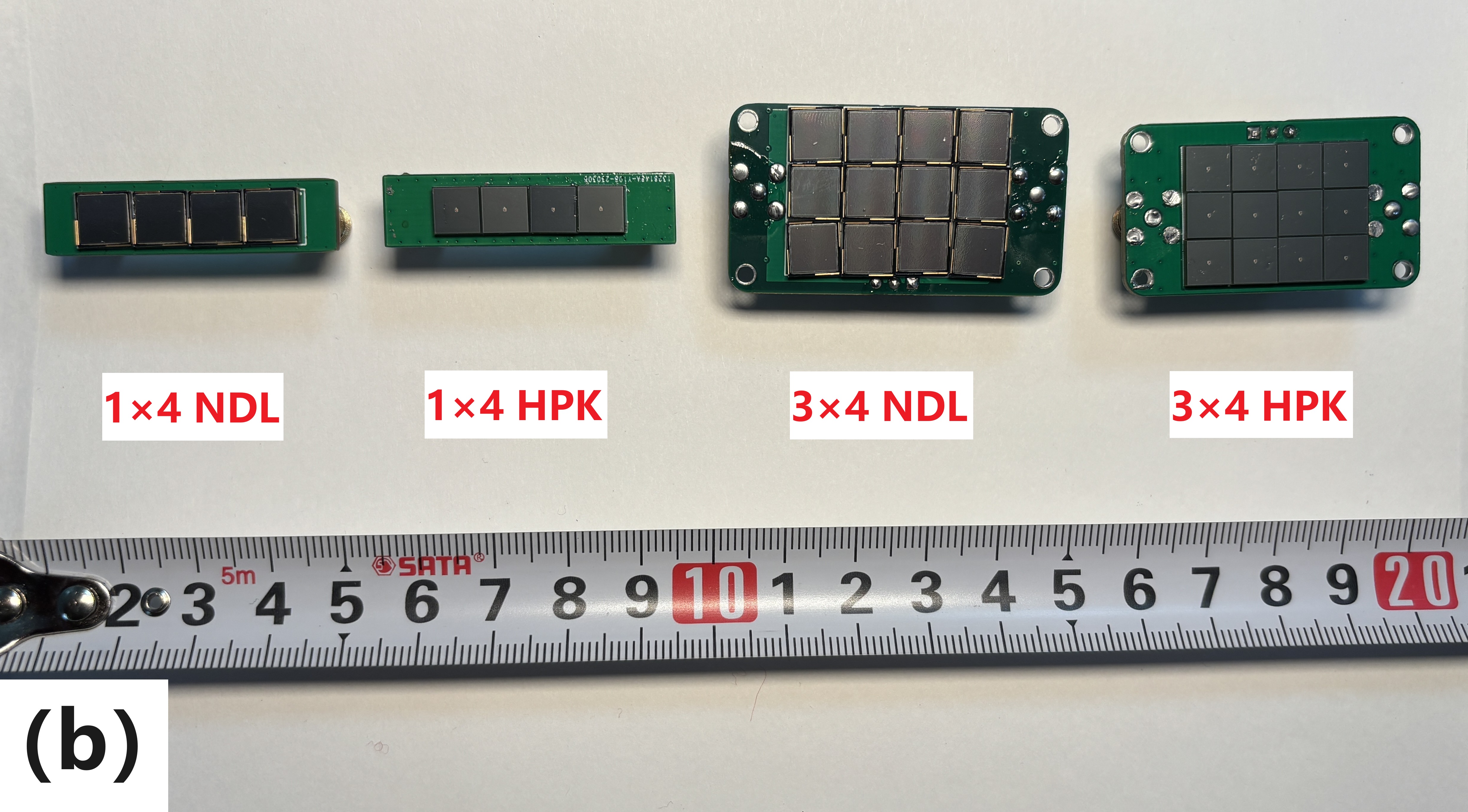}
\caption{(a) BC420 scintillator sample from Saint-Gobain alongside various scintillator samples from GaoNengKeDi. (b) Different configurations of SiPM arrays with a preamplifier mounted on the back of the PCB. }
\label{scintillator}
\end{figure}

In recent years, with the motivation for the R\&D of the Belle II KLM upgrade, our laboratory at Fudan University has been working closely with the GaoNengKeDi Company to improve the quality of scintillators, especially in terms of light yield and Bulk Light Attenuation Length ($\Lat$) ~\cite{GNKD}. Several scintillator samples were fabricated for experimental evaluation, utilizing high-transparency polystyrene and a light-emitting material with a decay time of approximately $1.5~\ns$. Our development process revealed that refining the manufacturing procedure is essential for improving the attenuation length of the plastic scintillator. As illustrated in Fig.~\ref{scintillator}(a), we catalog the GaoNengKeDi scintillators with dimensions ranging from $4~\cm \times 2~\cm \times 50~\cm$ to $4~\cm \times 2~\cm \times 150~\cm$ by the shipment date of the samples: GNKD\_1, GNKD\_2, GNKD\_3, and GNKD\_4.

The BC420 scintillator from Saint-Gobain~\cite{SG}, known for its benchmark quality in TOF detectors, served as a reference for evaluating cost-effective alternatives from GaoNengKeDi. The dimension of the BC420 scintillator is $5~\cm \times 3~\cm \times 120~\cm$, with a $\Lat$ of $110~\cm$~\cite{SG}. The BC420 scintillator is also depicted in Fig.~\ref{scintillator}(a). More information about the scintillator samples can be found in Table~\ref{tab}.

\subsection{SiPMs and front-end electronics}

All the SiPMs used in this study measure $6~\mm \times 6~\mm$, including the HPK S14160-6050~\cite{HPK} and the EQR20-11-6060 from Novel Device Laboratory (NDL, now known as Capital Photonics Inc. of the China General Nuclear)~\cite{NDL}. The HPK S14160 exhibits a lower dark count rate (DCR), while the EQR20, a newly developed SiPM, demonstrates significant advantages for large-scale detectors, such as an upgraded KLM. Unlike conventional SiPMs that rely on polysilicon resistors for quenching, the EQR design leverages the intrinsic resistance of the silicon epitaxial layer for avalanche quenching. This innovative approach increases microcell density and fill factor, achieving an optimal balance between high photon detection efficiency and a wide dynamic range. Additionally, it eliminates the need for silicon strip resistor fabrication, simplifying both device structure and manufacturing. The EQR SiPM also features low terminal capacitance, narrow pulse waveforms, and fast response times. Similar to the scintillator from GaoNengKeDi, the EQR SiPM from Capital Photonic offers a significant cost advantage.

Front-end electronics is essential for HTR measurements using a SiPM array.  We designed the preamplifier~\cite{PreAMP} using the ultra-low noise, high bandwidth operational amplifier LMH6629 from Texas Instruments~\cite{LMH6629}. This preamplifier provides a gain of +26 dB and a bandwidth of $426~\mhz$, enabling a $\Trise$ of $1~\ns$ while reducing the baseline noise level to as low as $0.6~\mV$. With photoelectron counts exceeding 60, the amplifier integrated with one SiPM achieves a time resolution of $20~\ps$~\cite{PreAMP}.

To further enhance the time resolution of a scintillator detector by increasing the photon collection area, we configure the multiple SiPMs into series, parallel, or series and parallel arrangements, forming SiPM arrays~\cite{SP}, as illustrated in Fig.~\ref{scintillator}(b). The preamplifier and SiPM array are mounted on the same printed circuit board (PCB) to minimize noise and interference during signal transmission.

\section{Experimental setup for cosmic ray measurements}
\label{sec:III}

\begin{figure}[htb]
\centering
\includegraphics[width=0.75\textwidth]{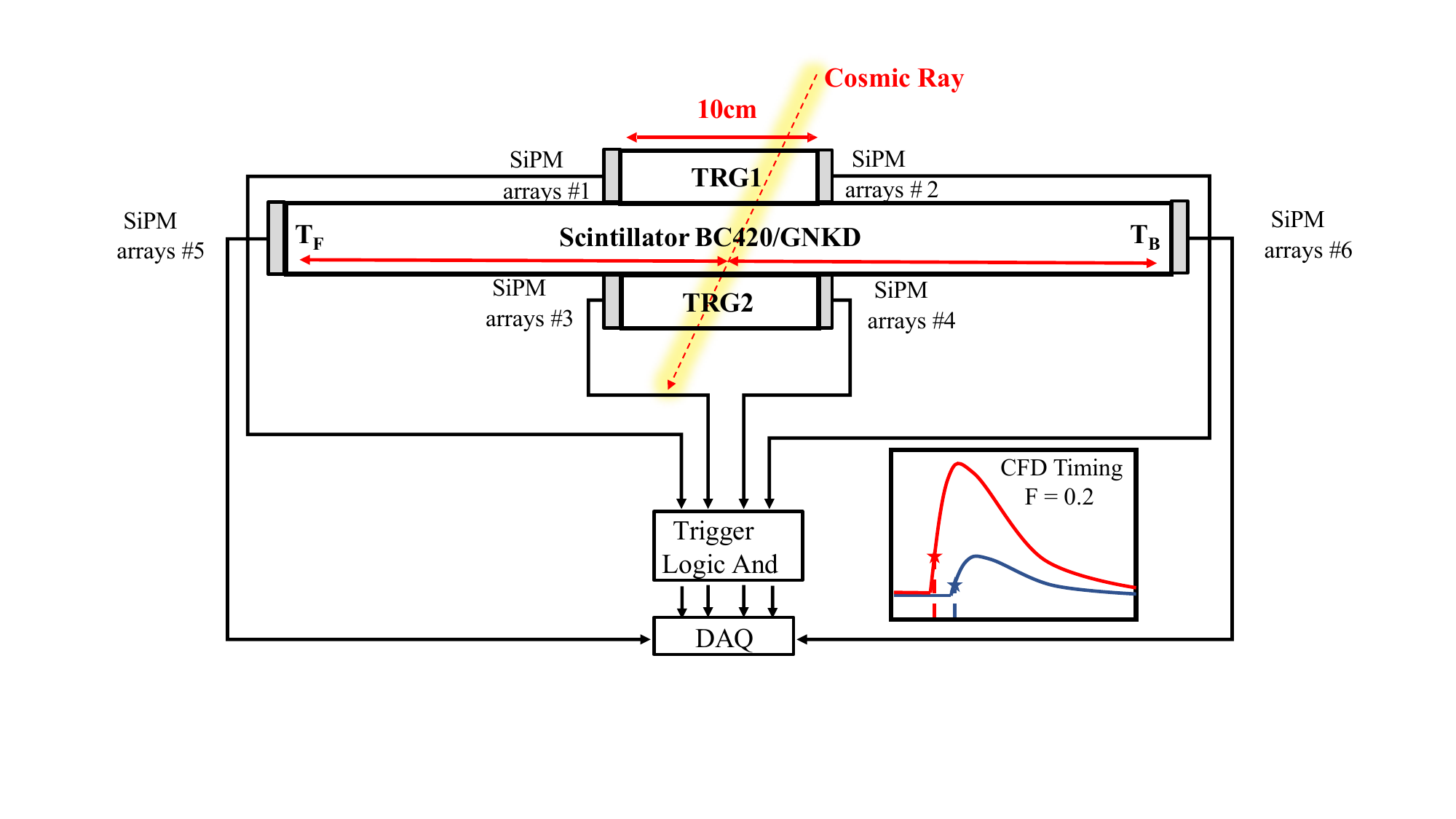}
\caption{Experimental setup for cosmic ray test on the long scintillators. }
\label{cr_setup}
\end{figure}

The performance of the scintillators is evaluated using cosmic ray (CR) testing. This approach simulates the performance of a detector channel in the upgraded KLM system, specifically in detecting hits generated by muons or pions within a hadronic cluster. Figure~\ref{cr_setup} shows the setup containing a trigger system, long scintillator(s) from GaoNengKeDi or BC420 under test, SiPM arrays for photon detection, and a data acquisition system (DAQ).

We employed the Tektronix MSO58 oscilloscope with eight channels or the CAEN desktop digitizer DT5742~\cite{DT5742} with 16 channels for the DAQ system. The oscilloscope (digitizer) features a sampling rate of 6.25 GS/$\s$ (5 GS/$\s$) and an analog bandwidth of $1~\ghz$ ($500~\mhz$). The resulting data is processed offline on a computer after digitizing the waveform via DAQ. To obtain accurate timing information, we apply the Constant Fraction Discrimination method to the waveform, with the fraction set to $f = 0.2$.

The trigger system consists of two $1~\cm \times 4~\cm \times 10~\cm$ plastic scintillator strips from GaoNengKeDi and a logic board for DAQ trigger signal input. Four $1\times 4$ SiPM arrays are mounted on the ends of the scintillator strips, with $T_i$ ($i=1,2,3,4$) representing the recorded time. The hit time in each strip is calculated as $T_{\rm trg1} = (T_1 + T_2)/2$ or $T_{\rm trg2} = (T_3+T_4)/2$, and the trigger time is determined by  $T_0 = (T_{\rm trg1} + T_{\rm trg2})/2$. To obtain the resolution of trigger time $\sigma_{T_0}$, we analyze the time difference $\Delta T_{\rm trg} = T_{\rm trg1} - T_{\rm trg2}$. The distribution of $\Delta T_{\rm trg}$ is fitted with a Gaussian function to determine its standard deviation $\sigma_{\Delta T_{\rm trg}}$, from which we derive $\sigma_{T_0} = \sigma_{\Delta T_{\rm trg}}/2$ and the time resolution of single readout $\sigma_{T_i} \approx \sigma_{\Delta T_{\rm trg}}$.

For a long strip under test, two $3\times 4$ SiPM arrays are mounted on the two ends, with their recorded times denoted as $T_F$ and $T_B$. The hit time on the long strip is calculated as $T_{\rm SC} = (T_F + T_B)/2$. To determine the time resolution of this detector, we analyze the time difference $\Delta T = T_{\rm SC} - T_0$. The distribution of $\Delta T$ is fitted with a Gaussian function to obtain its standard deviation $\sigma_{\Delta T}$. By subtracting the uncertainty of $T_0$, the time resolution of the long scintillator detector (LSD) is calculated via:
\beq
\label{eq_tof}
\sigma_{T_{\rm LSD}} = \sqrt{\sigma^2_{\Delta T} - \sigma^2_{T_0}}.
\eeq

\section{Time resolutions of SiPM arrays}
\label{sec:IV}

The $1\times 4$ and $3\times 4$ SiPM arrays in the trigger system and the LSDs are depicted in Fig.~\ref{scintillator}(b). We evaluate their time resolutions under different connection configurations: $1\times 4$ arrays in parallel (4P) and series (4S), and $3\times 4$ arrays in pure series (12S), pure parallel (12P), and hybrid (4S3P).  Parallel configurations increase the amplifier’s input junction capacitance, resulting in larger $\Trise$ and higher baseline noise. In contrast, series configurations effectively mitigate these issues and significantly shorten the $\Trise$ of the SiPM signal waveform, thereby improving the time resolution in CFD. For $1\times 4$ arrays, $\Trise$ improves from $6.5~\ns$ (4P) to $3.3~\ns$ (4S). For $3\times 4$ arrays, $\Trise$ improves from $12.4~\ns$ (12P) to $4.1~\ns$ (4S3P). Nevertheless, the extended signal path in the pure series configuration for large SiPM arrays may degrade time resolution. The 12S configuration offers no significant $\Trise$ improvement, decreasing only marginally from $4.1~\ns$ (4S3P) to $4.0~\ns$ (12S). Furthermore, the reduction in the junction capacitance decreases the output charge, which is reflected in a narrower pulse width while the pulse amplitude remains almost the same. Consequently, the 4S3P hybrid configuration emerges as the optimized choice for the $3\times 4$ SiPM arrays.

\begin{figure}[htb]
\centering
\includegraphics[width=0.45\textwidth]{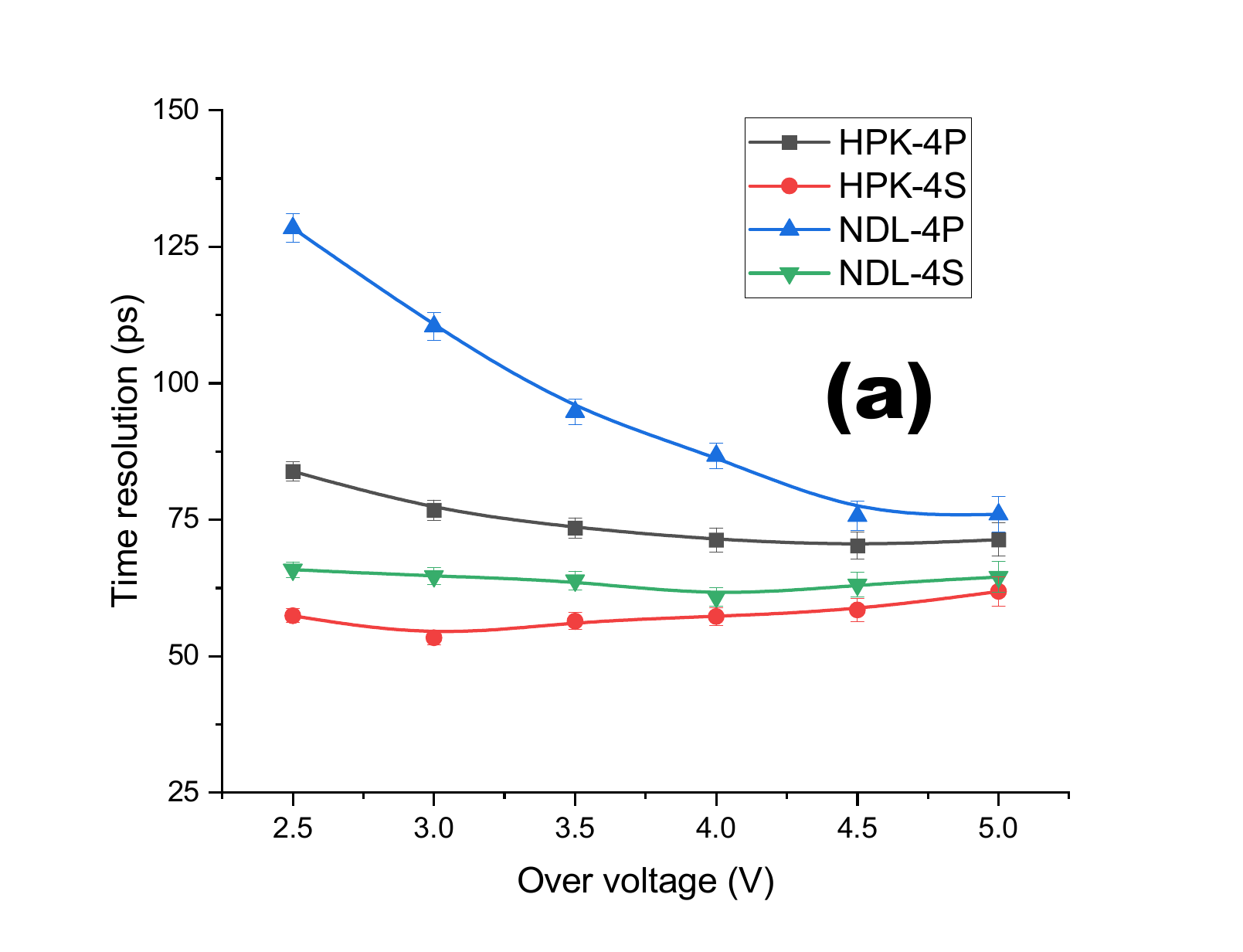}
\includegraphics[width=0.45\textwidth]{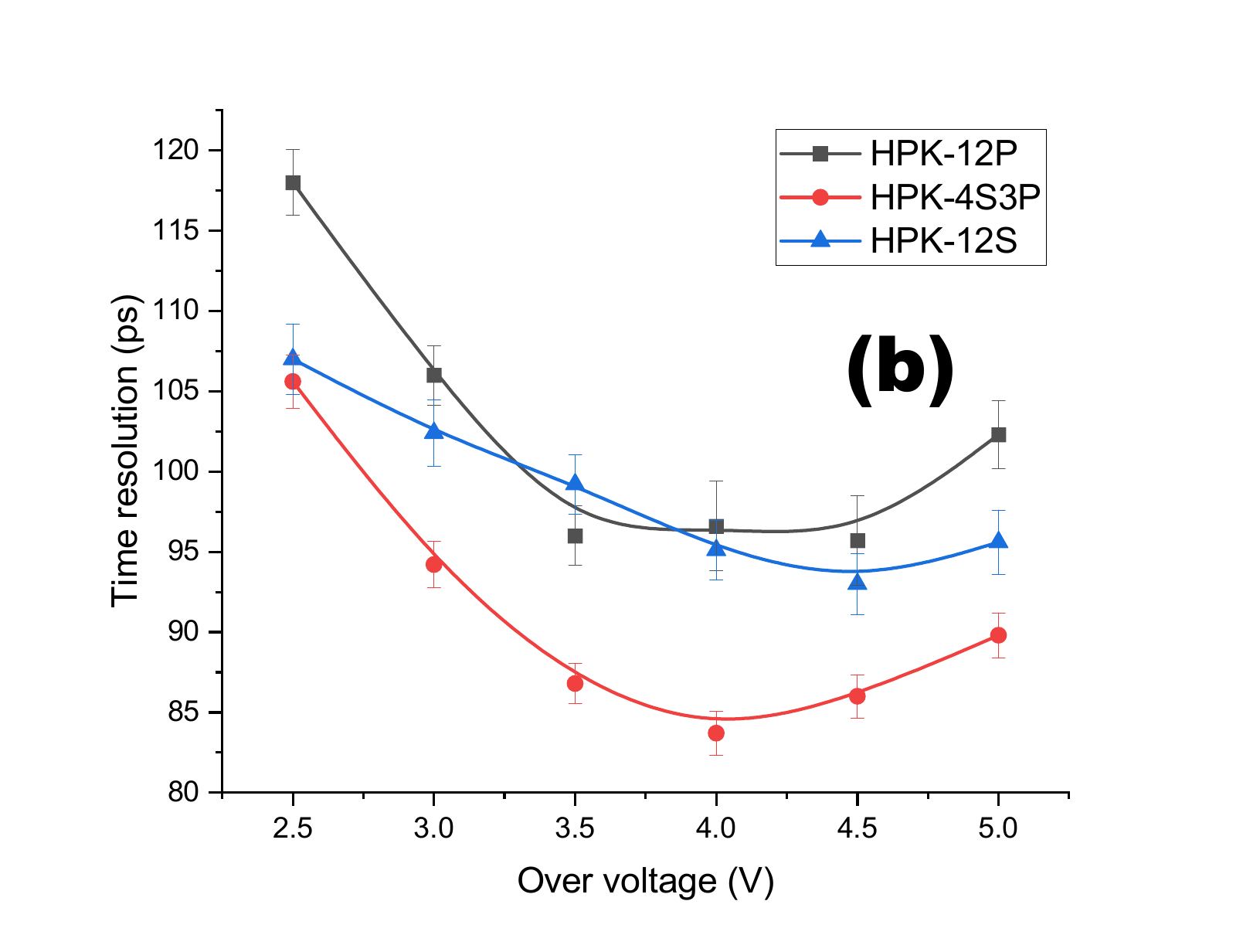}
\caption{Comparison of configurations for SiPM arrays in (a) the trigger system and (b) the detector using BC420 scintillator and their impact on time resolution as a function of overvoltage.}
\label{trg_reso}
\end{figure}

The overvoltage of the SiPM significantly impacts time resolution.  Figure~\ref{trg_reso}(a) shows the performance of $1\times 4$ SiPM arrays, revealing that the 4S configuration achieved superior time resolution compared to the 4P configuration. Notably, the HPK S14160-6050 outperformed the EQR20 SiPM from NDL in this regard, benefiting from the lower DCR. At the optimal overvoltage, the HPK 4S configuration yielded a time resolution of $\sigma_{\Delta T_{\rm trg}} = 104 \pm 4~\ps$. The 4P and 4S configurations of HPK SiPMs, as well as the 4S configuration of NDL SiPMs, demonstrate stable time resolution across varying overvoltages. However, the time resolution of the 4P configuration of NDL SiPMs degrades significantly when the overvoltage is reduced from $5~\V$ to $2.5~\V$. We choose the HPK 4S for the default trigger system, which has a time resolution of $\sigma_{T_0} = 52 \pm 2~\ps$. Figure~\ref{trg_reso}(b) presents the results for the $3\times 4$ SiPM arrays, indicating that the hybrid 4S3P configuration achieved the best time resolution, with an optimal overvoltage of approximately $4~\V$.

\section{Performance with single-end readout}
\label{sec:IV-B}

\begin{figure}[htb]
\centering
\includegraphics[width=0.55\textwidth]{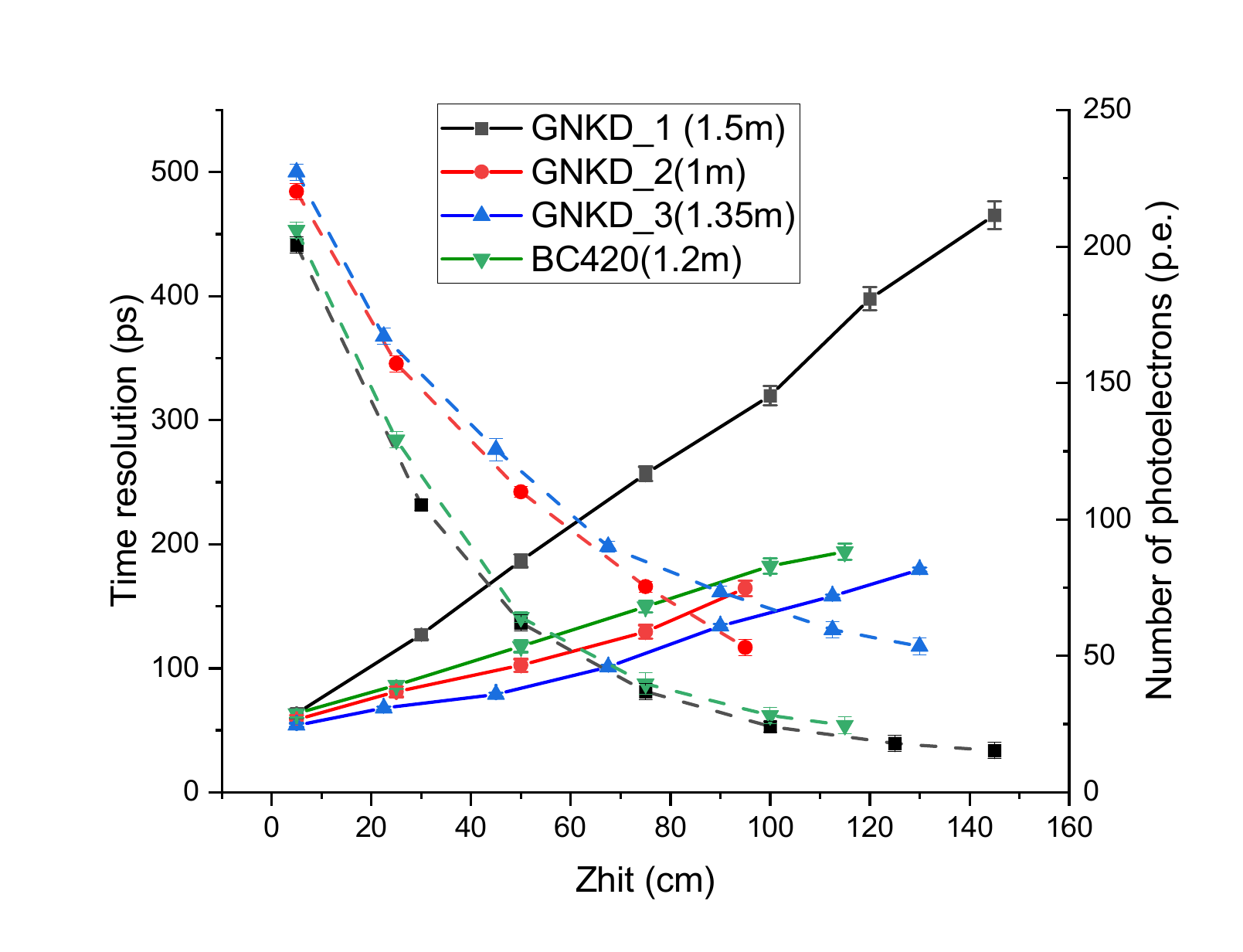}
\caption{The variation of photon collection and single-ended time resolution is analyzed in relation to the position of cosmic ray hits. The photon collection and time resolution are represented in dashed lines and solid lines,  respectively.}
\label{GNKD_NPE}
\end{figure}

By adjusting the positions of the upper and lower trigger trips, we can evaluate the average number of collected photoelectrons (nPE) and time resolution from single-end readout for cosmic rays hitting at various locations along the scintillator, as shown in Fig.~\ref{GNKD_NPE}.  The BC420 scintillator exhibits an nPE of approximately 460 at the near end, which decreases to 25 at the far end. The GNKD\_1 demonstrates comparable nPE performance, while the subsequently developed GNKD\_2 and GNKD\_3 show significant improvements. Notably, the $1.35~\m$ GNKD\_3 sample achieves an nPE of 500 at the near end and maintains 50 at the far end. These results indicate that GaoNengKeDi has successfully enhanced the transparency of their scintillators, fulfilling our R\&D requirements for the KLM upgrade.

The $\Lat$s of the scintillator bars are determined by performing exponential fits to the nPE values along with their lengths, and the results are listed in Table~\ref{tab}. Due to the effect of light reflection, the scintillator attenuation length measured in the laboratory is generally smaller than the intrinsic attenuation length. To ensure the measured $\Lat$ reflects the intrinsic attenuation length as accurately as possible, the influence of reflected light, which is more pronounced near the SiPM readout ends, is minimized by selecting data points located more than $50~\cm$ from the readout end for fitting. Fitting the data of BC420 yields $\Lat = 80\pm 7~\cm$. GNKD\_1 exhibits a similar nPE performance and a fitted $\Lat = 73\pm 7~\cm$, as shown in Fig.~\ref{GNKD_NPE}. GNKD\_2 has a shorter $\Lat = 63 \pm 2~\cm$, but its nPE is approximately 50\% higher at a distance of $50~\cm$ from the SiPM array. GNKD\_3 demonstrates superior nPE performance and a significantly larger attenuation length of $\Lat =115 \pm 5~\cm$ compared to all other samples.  $\Lat$ of GNKD\_4 is not measured due to its limited length of $50~\cm$. 

Unlike dual-end readout, single-end readout requires consideration of the positional uncertainty in cosmic ray hit locations when determining the time resolution. For instance, when calculating the timing resolution of  $T_F$ for the SiPM array\#5 in Fig.~\ref{cr_setup}, we utilize the left side of the trigger system, defined as $T_{\rm trgL} = (T_1 + T_3)/2$. The time difference $\Delta T_{\rm single} = T_F - T_{\rm trgL}$ mitigates the positional uncertainty. By fitting the distribution of $\Delta T_{\rm single}$ with a Gaussian function, we obtain its standard deviation $\sigma_{\Delta T_{\rm single}}$. Consequently, the time resolution of $T_F$ is calculated as:
\beq\label{eq_single_end}
\sigma_{T_F} = \sqrt{\sigma_{\Delta T_{\rm single}}^2 - \sigma^2_{T_i}/2}.
\eeq

\begin{table*}[htbp]
\centering
\caption{Comparison of photon collection and time resolution at the near-end and far-end of scintillators with different $\Lat$.}
\small
\label{tab}
\begin{tabular}{ l  c  c  c  c  c  c } 
\hline 
\multirow{2}{*}{Type} & \multirow{2}{*}{$W \times H \times L$ ($\cm^3$)} & \multirow{2}{*}{$\Lat$
($\cm$)} & \multicolumn{2}{c}{Time Resolution ($\ps$)} &  \multicolumn{2}{c}{nPE} \\
& & & Zhit=$5~\cm$ & $Z_{\rm hit} = 95~\cm$ & $Z_{\rm hit} = 5~\cm$ & $Z_{\rm hit} = 95~\cm$\\
\hline
SG-BC420& $5 \times 3 \times 120$ & $80 \pm 7 $ & $63 \pm 4$ & $175 \pm 6$ & 206 & 30\\
GNKD\_1 & $4 \times 2 \times 150$ & $73 \pm 7 $ & $63 \pm 4$ & $307 \pm 8$ & 200 & 26\\
GNKD\_2 & $4 \times 2 \times 100$ & $63 \pm 2 $ & $59 \pm 4$ & $164 \pm 6$ & 220 & 53\\
GNKD\_3 & $4 \times 2 \times 135$ & $115 \pm 5$ & $53 \pm 2$ & $138 \pm 6$ & 227 & 70\\
GNKD\_4 & $4 \times 2 \times 50 $ & 	/ 	  	  & $44 \pm 3$ & $81 \pm 5 ($45~\cm$) $ & 341 & 154 ($45~\cm$)\\
\hline
\end{tabular}
\end{table*}

The time resolution strongly correlates with nPE, as summarized in Table~\ref{tab}. All scintillators achieve excellent time resolutions of approximately $60~\ps$ at the near end. As the hit position moves farther from the readout end, scintillators with larger $\Lat$ and higher light yields, such as GNKD\_3, exhibit superior time resolution. Despite this advantage, the time resolution of GNKD\_3 degrades from $53 \pm 2~\ps$ at the near end to $140 \pm 6~\ps$ at the far end. These results underscore the necessity of implementing a dual-end readout scheme to maintain consistent performance across the entire length of the scintillator.

\section{Performance with dual-end readout}
\label{sec:IV-C}

The dual-end scheme averages the time information from both ends of an LSD, incorporating a weighting factor based on nPE. This approach is analogous to the method employed in traditional TOF detectors utilizing PMTs~\cite{TOF2}. The resolution of $T_{\rm SC}$ is determined by the weighted average of $T_{\rm F}$ and $T_{\rm B}$:
\beqar\label{eq_1}
T_{\rm SC} & = & \frac{T_{\rm F} \sigma_{T_F}^{-2} + T_{\rm B} \sigma_{T_B}^{-2}}
{\sigma_{T_{\rm SC}}^{-2}},
\eeqar
\beqar\label{eq_2}
\sigma_{T_{\rm SC}}^{-2} & =& \sigma_{T_F}^{-2} + \sigma_{T_B}^{-2},
\eeqar
with $T_F = T_0 + X/v_{\rm eff}$ and $T_B = T_0 + (L-X)/v_{\rm eff}$, where $X$ is the distance from the cosmic ray hit position to the readout end, $v_{\rm eff}$ is the effective propagation speed of light within the scintillator, and $\sigma_{T_F}$ , $\sigma_{T_B}$ represent the time resolution of the front-end and back-end readout of the plastic scintillator. 

\begin{figure}[htb]
\centering
\includegraphics[width=0.45\textwidth]{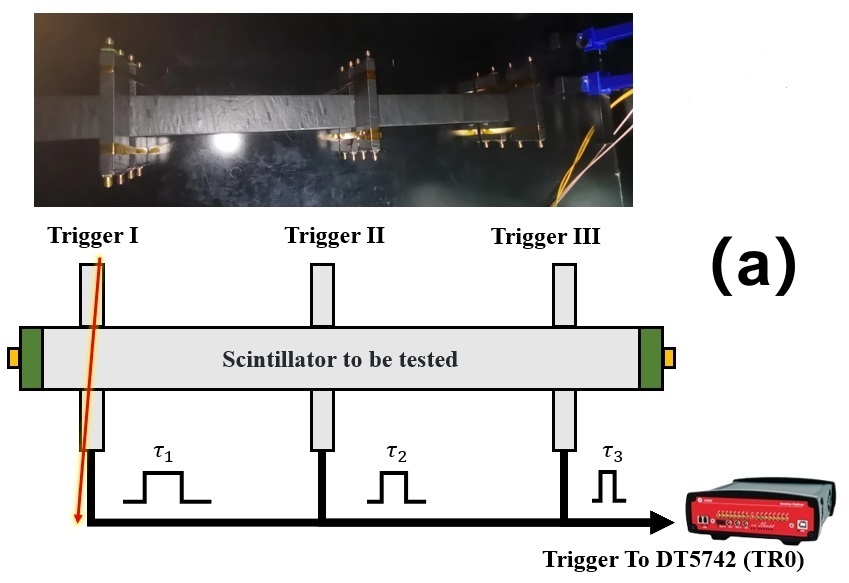}
\includegraphics[width=0.45\textwidth]{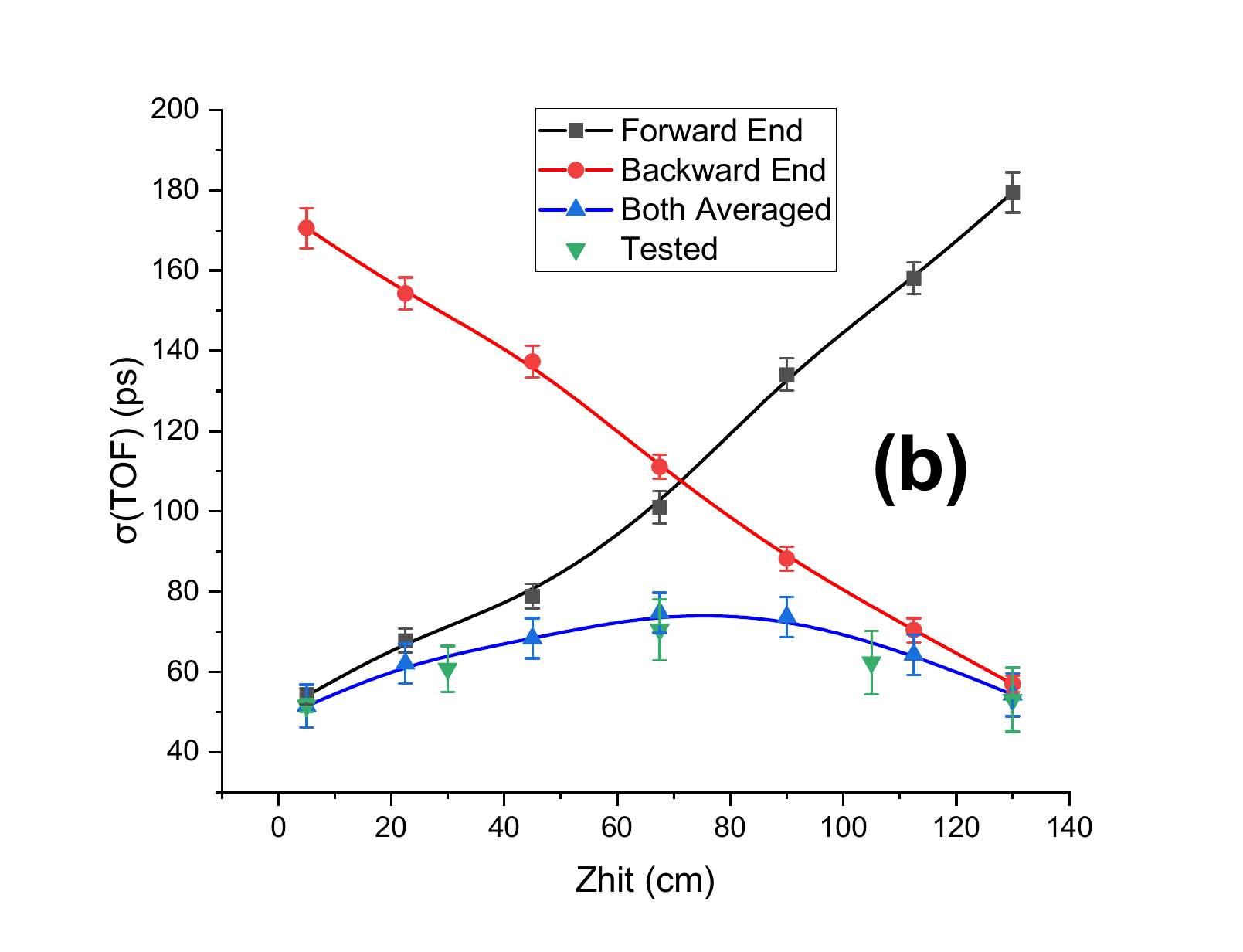}
\caption{Time resolution of the GNKD\_3 sample measured in cosmic ray test. (a) Schematic diagram of the multi-trigger test system. (b) Time resolution as a function of scintillator position. The red and black lines represent single-end time resolution, while the blue line corresponds to calculations using Eq.~(\ref{eq_1}). The green line represents the time resolution results obtained by averaging the times at both ends, with the trigger length reduced to $1~\cm$.}
\label{TOF_Zhit}
\end{figure}

The precision of $T_{\rm SC}$ is affected by uncertainties in the muon hit position after weighted averaging, except near the scintillator midpoint, where both ends contribute equally. Although a short $10~\cm$ scintillator is used as the trigger, this length still introduces considerable position uncertainty. Reducing the trigger length to $1~\cm$ would mitigate this, but also decrease cosmic ray flux, requiring several weeks per position. To improve efficiency, we propose a multi-trigger scheme with trigger pulses of different widths ($\tau_1 = 50~\ns$, $\tau_2 = 100~\ns$, $\tau_3 = 150~\ns$) fed into the DT5742's TR0 interface, as illustrated in Fig.~\ref{TOF_Zhit}(a). These distinct pulses allow cosmic ray hit locations to be determined during analysis, enabling simultaneous measurement at multiple positions and significantly enhancing test efficiency.

We conducted CR tests on GNKD\_3 to further evaluate their superior time resolution. Figure~\ref{TOF_Zhit}(b) compares the dual-end readout and single-end readout schemes. The poorest resolution,  $70 \pm 9~\ps$, is observed at the midpoint, while the optimal resolutions of approximately $50~\ps$ are achieved at both ends. The red and black curves represent the time resolutions of the two ends in the single-end readout scheme. The blue curve shows the calculated values based on Eq.~(\ref{eq_1}), which align well with the direct measurements. This calculation can be extended to a scintillator with a length of $200~\cm$, the typical length of a long scintillator strip in the KLM detector, yielding a time resolution of $98~\ps$ at its midpoint. Using the dual-end readout, GNKD\_3 demonstrates timing performance that meets the requirements for the KLM upgrade, specifically for measuring the $T_{\rm fly}$ of a $\kl$ meson~\cite{B2U_CDR}.  Meanwhile, the $50~\cm$ GNKD\_4 demonstrates a time resolution of $44 \pm 3~\ps$ at the near end, $81 \pm 5~\ps$ at the far end, and $47 \pm 2~\ps$ at the midpoint when utilizing a dual-end readout configuration.

\section{Measurement of cosmic ray velocity}
\label{sec:IV-D}

An upgraded KLM should incorporate multiple layers to facilitate the reconstruction of hadronic showers from $\kl$ mesons or to track multiple hits from a muon track~\cite{B2U_CDR}. To simulate this application, we assembled an array using GaoNengKeDi scintillators to measure the average velocity of cosmic rays. Due to the limited number of scintillator samples in our laboratory, we opted for $75~\cm$ long scintillators from batch GNKD\_1, which exhibited a consistent time resolution of $120 \pm 5~\ps$, for constructing the prototype array.

\begin{figure}[htb]
\centering
\includegraphics[width=0.55\textwidth]{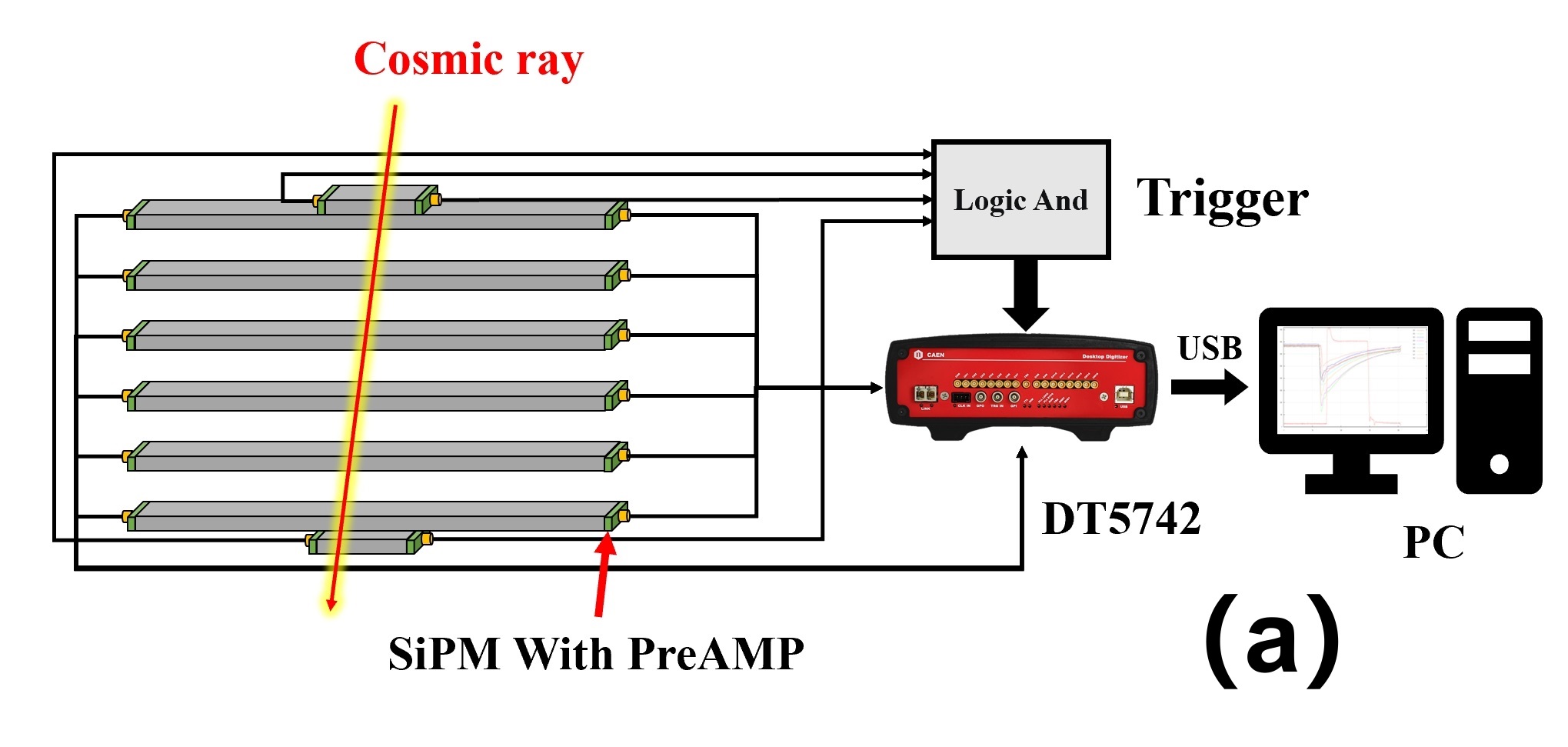}\\
\includegraphics[width=0.45\textwidth]{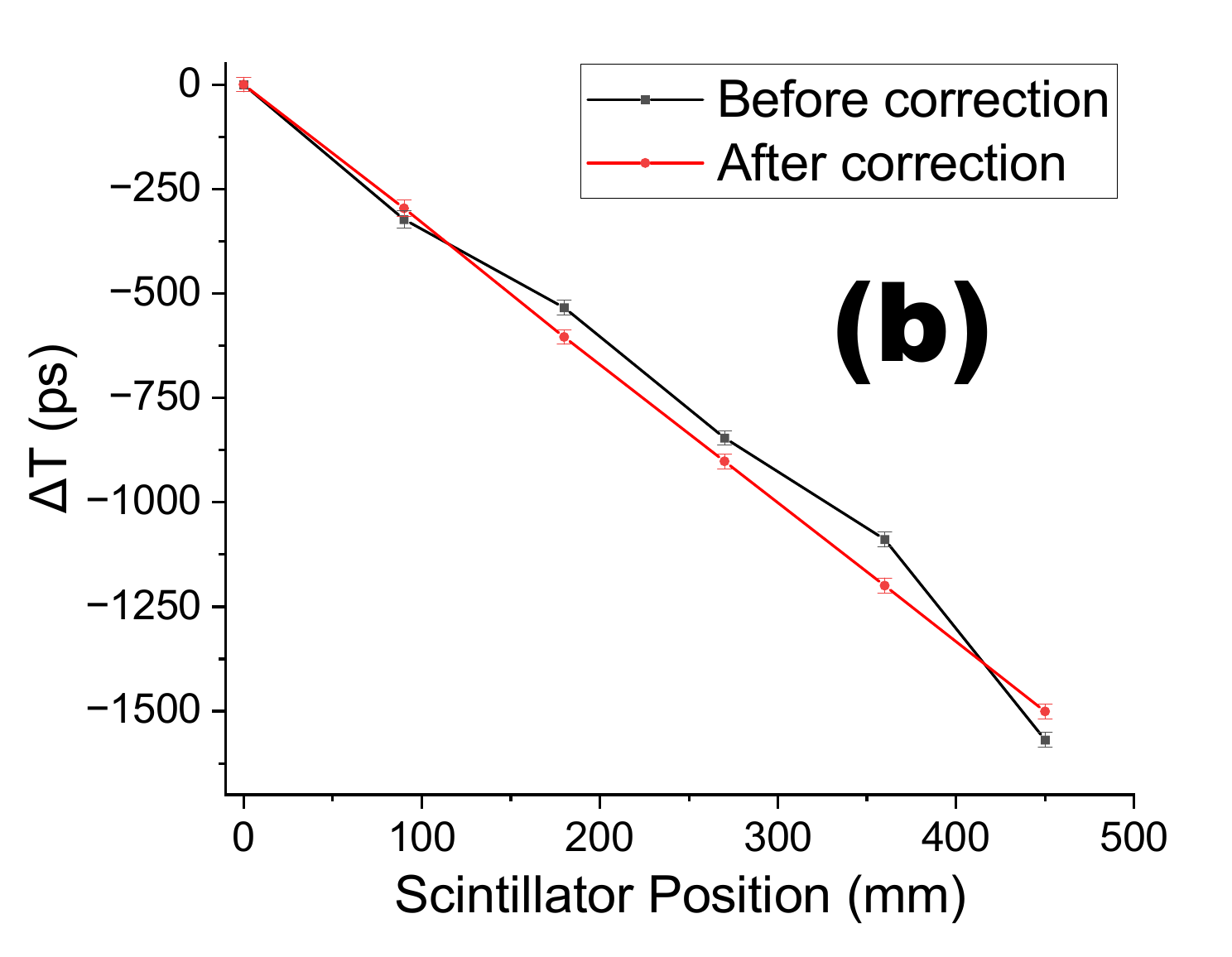}
\includegraphics[width=0.45\textwidth]{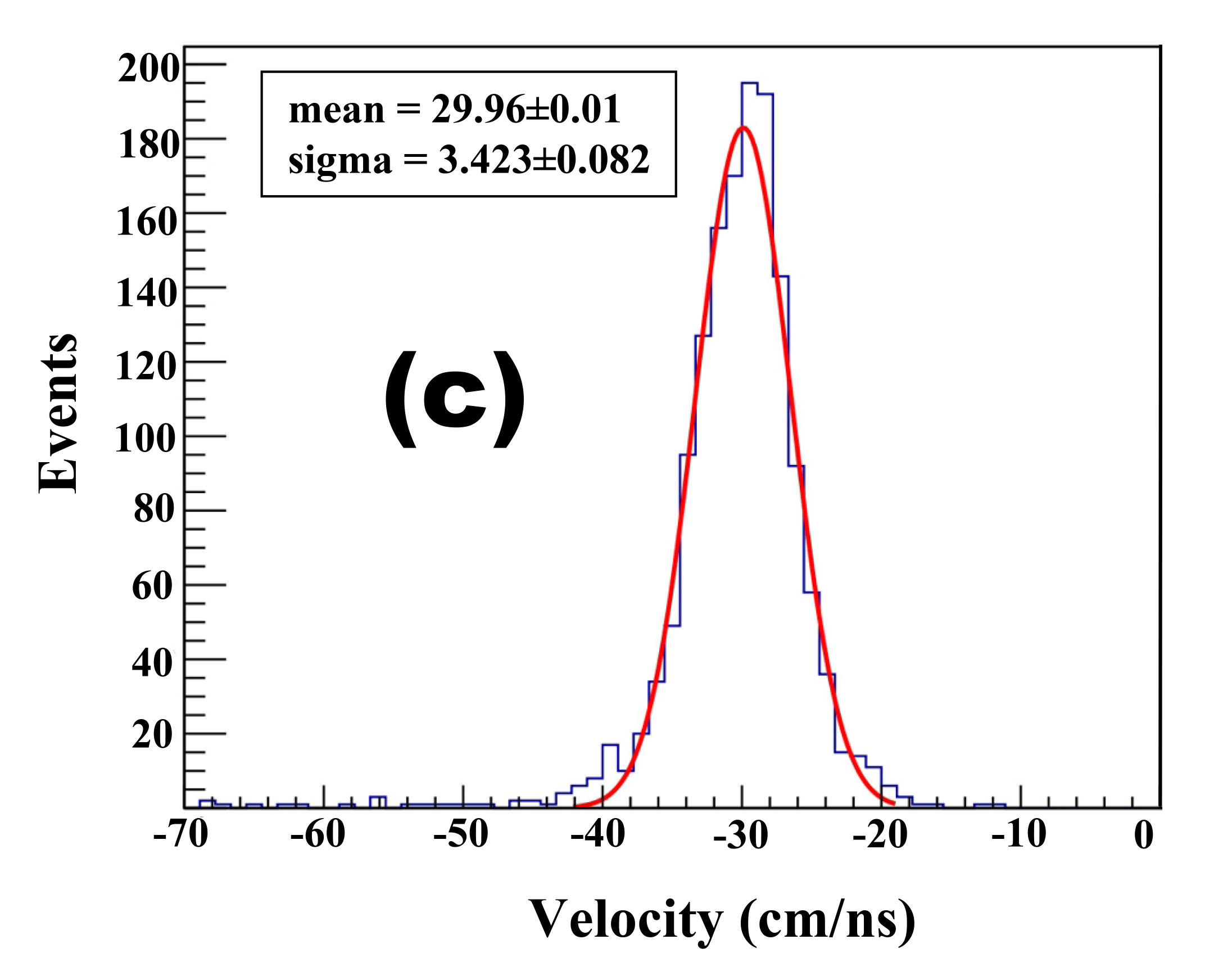}
\caption{(a) Schematic of the prototype setup, featuring six long scintillators measuring $75~\cm \times 4~\cm \times 1.5~\cm$ and two short trigger scintillators measuring $10~\cm \times 4~\cm \times 1~\cm$. The distance between each pair of long scintillators is $7.5~\cm$. (b) Timing versus the position of the long scintillators for cosmic ray velocity measurements, comparing results before and after time calibration of the readout channels. (c) The Gaussian fit to the velocity distribution of all measured cosmic ray hits.}
\label{prototype}
\end{figure}

The prototype apparatus is illustrated in Fig.~\ref{prototype}(a). It comprises six parallel $75~\cm$ scintillator strips arranged vertically with an equal spacing of $7.5~\cm$, labeled from top to bottom as SC1, SC2, SC3, SC4, SC5, and SC6. Two trigger strips are positioned at the top and bottom of the central region of the long strips. Dual-end readout is implemented for all eight strips to ensure precise timing measurements. Waveform acquisition is performed using the DT5742 digitizer, where the four channels of the trigger system generate logic signals that are routed into the TR0.

As shown in Fig.~\ref{prototype}(b), the timing versus scintillator position for the six strips deviates from a straight line due to variations in the electronic readout channels. To address these variations, we perform a calibration using a pulsed laser. The laser is split into six beams, each directed to the midpoint of the corresponding scintillator. By analyzing the timing discrepancies in the readouts and setting $T_{\rm SC1} = 0~\ps$ for the first channel, the calibration constants for the other five channels are determined to be $\Delta T_{\rm SC2} = 27~\ps$, $\Delta T_{\rm SC3} = -70~\ps$, $\Delta T_{\rm SC4} = -56~\ps$, $\Delta T_{\rm SC5} = -111~\ps$, and $\Delta T_{\rm SC6} = 68~\ps$. After calibration, the timing versus scintillator position for the six strips aligns along a straight line, as shown in Fig.~\ref{prototype}(b).

We then perform a linear fit of the timing versus position to determine the velocity in each event. Fitting the distribution of velocities yields an average velocity of $29.958 \pm 0.011~\cm/\ns$, with a standard deviation of $3.4\pm 0.1~\cm/\ns$, as shown in Fig.~\ref{prototype}(c). This corresponds to $\beta = 0.9993$ and a momentum of $2.8 \pm 0.7~\gevc$ for muons in our laboratory, which is located on the third floor of an eight-floor building. The measured velocity aligns with the expected average muon momentum of $3 - 4~\gevc$ at the sea level in the Shanghai area~\cite{muon}. These results demonstrate the potential for improved HTR measurements with hits in multiple layers.

\section{Conclusion}

In summary, we present the R\&D efforts for the KLM upgrade proposal of the Belle II experiment, focusing on the feasibility of accurately determining the momentum of neutral hadrons, such as $\kl$ mesons and neutrons, through TOF measurements. We explored the use of newly developed cost-effective plastic scintillators with significant attenuation lengths, coupled with large-area SiPM arrays, to achieve excellent time resolutions. Notably, the $135~\cm$-long sample demonstrated a $\Lat$ of $120 \pm 7~\cm$ and a time resolution of $70\pm 7~\ps$ at its midpoint, and the latest $50~\cm$ scintillator achieved an exceptional time resolution of $47\pm 2~\ps$. To validate the applicability of this technology in the Belle II experiment, we constructed a prototype detector to measure the average cosmic ray velocity, obtaining a value of $29.958 \pm 0.011~\cm/\ns$, corresponding to a muon momentum of $2.8 \pm 0.7~\gevc$ in our laboratory. These results highlight the potential of direct momentum measurement of a neutral hadron in an upgraded Belle II KLM system.

\section*{Acknowledgment}

This work is partially supported by the National Key R\&D Program of China under Contract Nos. 2022YFA1601903 and 2024YFA1611002; the National Natural Science Foundation of China under Contract No. 12175041.

During the preparation of this work, the authors used DeepSeek and ChatGPT to improve language and readability. After using these tools, the authors reviewed and 
edited the content as needed and take full responsibility for the content of the publication.


\end{document}